# Effective, Practical PON Monitoring Beyond the Splitter

**Neil Parkin, Sophie Minoughan, Md Asif Iqbal**
*BT Research UK, neil.parkin@bt.com*

**Abstract:** Monitoring beyond the splitter in a PON is costly due to the need for additional hardware. A non-standard monitoring wavelength can reduce cost and increase the visibility of customers to 97% on a C+ GPON. © 2022 The Author(s)

## 1. Introduction and Motivation

Passive optical networks (PONs) are the network architecture of choice for residential fiber deployments. A PON is designed specifically to be cost-effective for delivering high data-rates to large customer populations. Unfortunately, this focus on low-cost data delivery limits traditional monitoring techniques. The optical power splitter creates a constraint due to its high loss and the overlapping of backscattered signals returned from multiple drop fibers, thus reducing the visibility of customer equipment. The conventional optical monitoring technique of Optical Time Domain Reflectometry (OTDR) has, so far, been unable to overcome these challenges. Numerous methods[1] have been proposed, but these require either active bypass elements such as optical switches[2] or passive bypass elements such as wavelength selective filters[3] and all additional elements come with substantial cost implications. The current default method is to use a U-band high dynamic range OTDR and a highly reflective device, such as a fiber Bragg grating or thin-film filter to increase detectability beyond the splitter. The Optical Network Terminal (ONT) within the customer home forms a significant post-splitter interest point. However, placing a reflector at this point is costly when scaled to millions of units, increases the inventory complexity, reduces the optical power budget and can be removed or damaged by the customer.

Ideally, an ONT reflector would not be required, the best scenario would be to confirm the ONT presence at the physical layer without any additional hardware. A wavelength that has a strong reflection by the ONT in question but is not functionally useful to operators is ideal. The water peak region centered at 1383 nm fits both these criteria and offers extra incentives. For example, higher wavelengths on G.657 bend-insensitive fibers, commonly deployed in the customer's residence become highly attenuated when bent. One 10 mm radius bend on a G.657.A1 fiber is allowed to add 1.5 dB at 1625 nm, thus reducing visibility of the ONT by the OTDR. At 1650 nm, as shown in Figure 1, this can approach 2 dB, but at 1383 nm the same bend adds less than 0.2 dB. This loss may be useful to identify a bend but not when trying to identify the ONT presence. Furthermore, given that the data transmission is largely unaffected by these bends, their identification is not especially desirable for fault detection

In terms of commercial viability, due to the high upstream dispersion of 1383 nm, it is unlikely to become a standardized PON transmission wavelength, where low-cost is imperative. In addition, there is historical reluctance to use this area due to the high attenuation that is dominant in older G.652.A/B fibers at 1383 nm due to OH- ion absorption, or the water-peak window (Figure 1). In reality, operators' fiber estates will be equal to or higher specification than the newer G.652.D and have comparable attenuation to 1310 nm. In fact, the amount of low water-peak fiber (G.652.D or similar) deployed has been reported as >80% globally between 1993 and 2018[4].

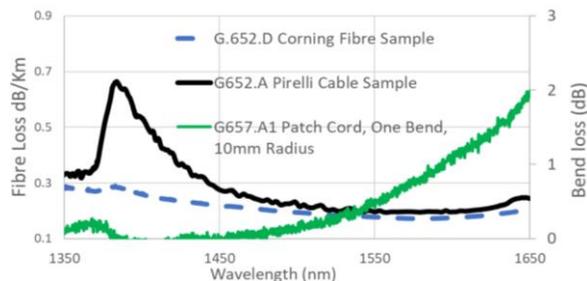
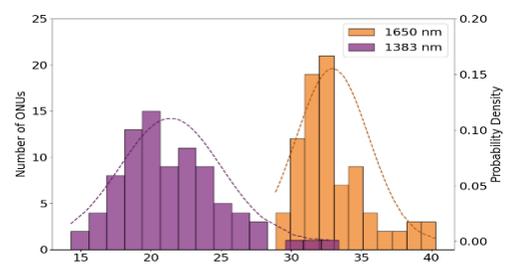

Figure 1: Measured attenuation and bend Loss    Figure 2: ORL and probability density function of ONTs measured

## 2. Tests, Results and Analysis of Benefits

The optical return loss (ORL) of eighty-seven GPON ONTs across four different vendors was measured (Figure 2). GPON ONTs were selected as this is the most deployed PON technology, with >51 million Optical Line Terminal (OLT) ports shipped vs 9.8 million for 10G PON since 2011[5]. Hence, GPON will require the most in-life live monitoring at a low cost. A continuous wavelength optical source and a 50/50 coupler was used to measure the ONT reflection. Figure 2 shows the measured ORL at the two wavelengths of interest with an 11.5 dB improvement in mean ORL at 1383 nm when compared with 1650 nm. This increased reflectance at 1383 nm has the potential to correlate with an increase in OTDR-based detection and thus, improve device visibility beyond the splitter compared with current OTDR wavelengths. This benefit relies on the reflectance of the filters used in the optical sub-assembly for coupling and separation of the upstream and downstream signals. At 1383 nm the reflectance is not specified or tested, and this gives a broader response due to a loose tolerance on the filters.

To evaluate the potential advantage, a 1383 nm OTDR with a datasheet dynamic range of 37 dB was used to assess a real-life network. A representative PON was set-up as shown in Figure 3, and the OTDR operated with a 30 ns pulse width and a 3-minute averaging time. This was compared to a 1650 nm OTDR designed for PON monitoring with a 46 dB dynamic range and the same pulse width and averaging. The first test used a Variable Back Reflector (VBR) to understand OTDR-sensitivity to a range of post-splitter reflections. A variable optical attenuator (VOA) was varied to adjust the attenuation to the standardized PON loss budgets of B+, C+ and D. The VBR was then adjusted to find a detectable reflection level at 1/1.5/2 dB above the measured noise floor for each instrument.

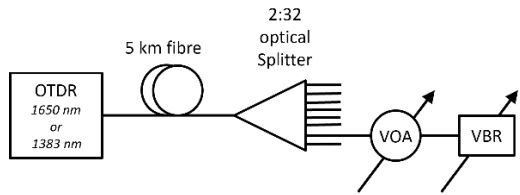
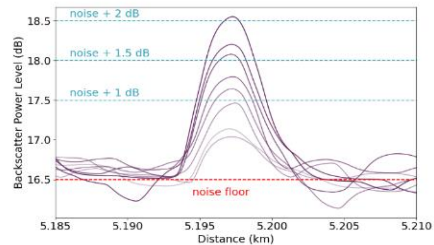

Figure 3: PON set-up for sensitivity measurements    Figure 4: Visibility beyond the splitter for C+ PON at 1383nm

The OTDR's sensitivity to the VBR reflectance is shown in Figure 5. This shows that the 1650 nm OTDR has better sensitivity as indicated by the datasheet dynamic range. This measurement can now be used to assess the proportion of ONTs that would be detectable at 1 dB above the OTDR noise floor. Using this data, and the ORL of the ONTs measured in Figure 2, the theoretically detectable ONT population was determined at both wavelengths and is shown in Figure 6. The increased reflection of the ONT allows for more units to be detected at 1383 nm irrespective of its inferior dynamic range.

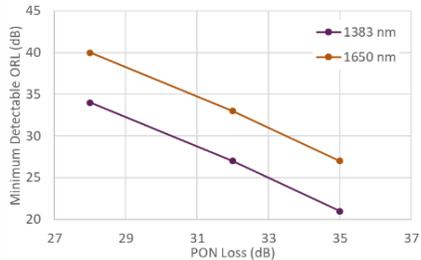
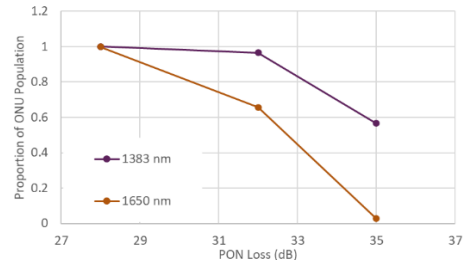

Figure 5: OTDR Sensitivity to VBR ORL    Figure 6: Proportion of ONTs detectable at PON budgets

A B+ PON, with a 1650 nm OTDR could detect 97% of the ONT population. This rises to 99% with 1383 nm. At C+, the proportion detected increases from 66% to 97% for 1650 and 1383 nm respectively and at the most challenging budget of 35 dB (D PON) where 1650 nm cannot detect anything, the 1383 nm wavelength OTDR can measure 57% of the ONTs.

Using an ONT sub-sample covering the range of ORLs, example traces were collected from the two OTDR devices. Figure 8 shows the range of ORL of the selected units, their respective detectability and where these sit within the ORL distribution for the complete ONT population. This confirms the advantage gained. Figure 7 shows example OTDR traces taken for a PON with a 20 km feeder and 32-way splitter for the same ONT on a C+ PON.

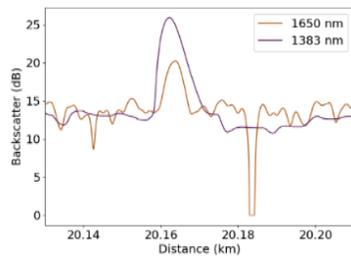
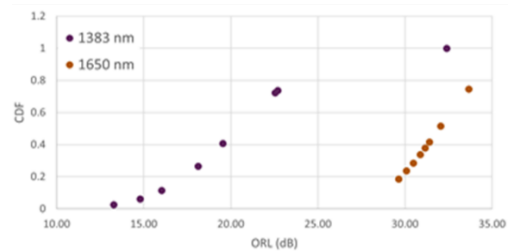

Figure 7: OTDR traces post-splitter at 20 km, C+ PON    Figure 8: ORL detected for the sample ONT population

## 3. System Considerations

The use of a non-standard wavelength in the optical distribution network requires an analysis of the impact it has on any service delivered on that network. In GPON the ITU has standardized the isolation to interference signals in G.984.5 to be >40 dB for the 1380 nm region. The impact was tested using a commercial GPON as shown in Figure 9. A signal of -7.8 dBm post-splitter would require no power penalty, allowing >+9 dBm at the headend with no impact on the customer. This is in line with the peak powers of current OTDR systems.

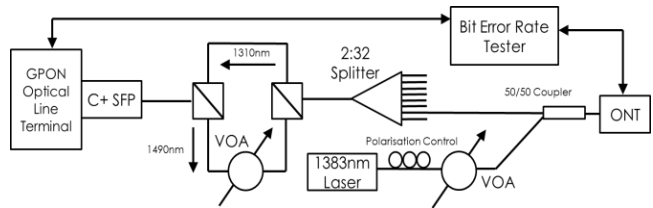
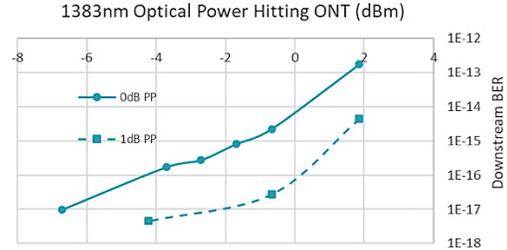

Figure 9: Experimental set-up for interference measurements    Figure 10: Impact of 1383 nm on system performance

Splitter attenuation at 1383 nm is not specified by the ITU (G.671; specified between 1260 to 1360 nm and 1480 to 1660 nm). Performance was measured on each port of three, 32-way planar lightwave circuit-based splitters. The average loss at 1383 nm was 15.76 dB with a uniformity of 0.69 dB. Compared to 1650 nm, which was 16.13 dB with a uniformity of 0.53 dB. Operating at 1383 nm gains a loss advantage over 1650 nm of 0.37 dB.

Another consideration is Stimulated Raman Scattering (SRS) induced power transfer between the OTDR channel and any downstream signal as the monitoring wavelength is now in the same region as transmission. In the case of current PON technologies, GPON's downstream 1490 nm signal is the worst at ~100 nm away, and HSPON at 1340-1344 nm at around 40 nm away. Both are susceptible as they are co-propagating signals. Numerical simulations were carried out using standard G.652.D attenuation values for a 22 km feeder fiber with a Raman gain coefficient for GPON of 0.3 ($W^{-1}km^{-1}$) and for HSPON as 0.17 ($W^{-1}km^{-1}$).

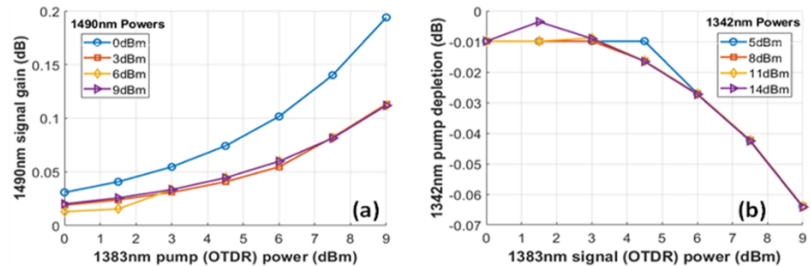

Figure 11: SRS induced (a) on-off Raman gain on 1490nm (GPON); and (b) depletion of HSPON (1342nm) signal

In GPON, the worst case is a low-data signal at 0 dBm, and a high OTDR signal at 9 dBm which amplifies the 1490 nm signal by ~0.2 dB (Figure 11a). As the OTDR acts as a pump it can be mitigated by reducing the OTDR power if the data signal is low. In HSPON (Figure 11b), the depletion of the 1342 nm signal by the OTDR channel is negligible (< 0.1 dB) at different powers due to lower Raman gain efficiency between those proximity channels. This depletion can be further reduced by lowering the OTDR channel power below 9 dBm.

## 4. Conclusion

An improvement over the standard monitoring wavelength in GPON is presented here. The gains are substantial and allow virtually full coverage in a C+ PON. This has a negligible effect on any existing deployed GPON systems. Newer PON systems would need to be assessed and, ideally, a set level of reflection at 1383 nm standardized.